\documentclass[onecolumn,10pt,a4paper,sort&compress,aps,pra,showpacs,superscriptaddress]{revtex4-2}

\usepackage{amsmath,amssymb,amsthm,mathtools,bm}
\usepackage{multirow}
\usepackage{hyperref}
\usepackage[capitalise,nameinlink]{cleveref}
\usepackage{graphicx}
\usepackage{overpic}
\usepackage{color}

\renewcommand{\vec}[1]{\bm{#1}}
\newcommand{\intd}{\mathop{}\!\mathrm{d}}

\newcommand{\diff}[2]{\frac{\mathrm{d}{#1}}{\mathrm{d}{#2}}}
\newcommand{\pdiff}[2]{\frac{\partial{#1}}{\partial{#2}}}

\newcommand{\bigO}[1]{\mathit{O}(#1)}

\newcommand{\avg}[1]{\overline{#1}}

\newcommand{\e}{\vec{e}}
\newcommand{\ex}{\e_x}
\newcommand{\ey}{\e_y}

\newcommand{\p}{\vec{p}}
\renewcommand{\d}{\vec{d}}

\begin{document}

\title{Systematic parameterisations of minimal models of microswimming}

\author{Benjamin J. Walker}
\email{benjamin.walker@ucl.ac.uk}
\affiliation{Department of Mathematics, University College London, London, WC1H 0AY, UK}

\author{Kenta Ishimoto}
\email{ishimoto@kurims.kyoto-u.ac.jp}
\affiliation{Research Institute for Mathematical Sciences, Kyoto University, Kyoto, 606-8502, Japan}

\author{Eamonn A. Gaffney}
\email{gaffney@maths.ox.ac.uk}
\affiliation{Mathematical Institute, University of Oxford, Oxford, OX2 6GG, UK}

\date{November 2022}


\begin{abstract}
Simple models are used throughout the physical sciences as a means of developing intuition, capturing phenomenology, and qualitatively reproducing observations. In studies of microswimming, simple force-dipole models are commonplace, arising generically as the leading-order, far-field descriptions of a range of complex biological and artificial swimmers. Though many of these swimmers are associated with intricate, time varying flow fields and changing shapes, we often turn to models with constant, averaged parameters for intuition, basic understanding, and back-of-the-envelope prediction. In this brief study, via an elementary multi-timescale analysis, we examine whether the standard use of \textit{a priori}-averaged parameters in minimal microswimmer models is justified, asking if their behavioural predictions qualitatively align with those of models that incorporate rapid temporal variation through simple extensions. In doing so, we highlight and exemplify how a straightforward asymptotic analysis of these non-autonomous models can result in effective, systematic parameterisations of minimal models of microswimming.
\end{abstract}

\maketitle

\section{Introduction}\label{sec: intro}
In the physical sciences, simple mathematical models are often used as a means of developing intuition and capturing phenomenology for complex systems. Whilst more complex, faithful models might require computational methods or advanced analytical techniques to study them, simple models are commonly amenable to succinct pen-and-paper analysis, frequently provide insight, and can sometimes yield dynamical predictions that agree with those of more intricate representations, at least qualitatively. This minimalistic approach is readily exemplified in the study of microswimmers, where minimal models of complex, shape deforming swimmers on the microscale are used in a range of settings, from back-of-the-envelope calculation and undergraduate teaching through to state-of-the-art, application-driven research contributions \cite{Zottl2012,Thery2020,Gutierrez-Ramos2018,Qi2022,Spagnolie2015,Elgeti2016,Malgaretti2017,Mathijssen2016,Lauga2020,Omori2022,Lauga2009}.

A common, if not ubiquitous, hydrodynamic representation of a microswimmer is the \emph{force dipole} model, wherein the flow field generated by a swimmer is taken to correspond simply to that of a force dipole. This approximate representation is valid in the far field of a microswimmer, with force-free swimming conditions being appropriate in the inertia-free limit of low-Reynolds-number swimming that applies to many microswimmers, including the well-studied spermatozoa and the breaststroke-swimming algae \textit{Chlamydomonas reinhardtii}. Invariably, the force dipole is assumed to be aligned along a swimmer-fixed axis and taken to be of constant signed strength. Swimmer parameters can be estimated from experimental measurements and hydrodynamic simulations \cite{Drescher2010, Guasto2010, Drescher2011,  Klindt2015, Ishimoto2017a, Ogawa2017, Ishimoto2020, Giuliani2021}, and typically involve averaging out rapid temporal variations that can be present in biological swimmers. This leads to a minimal model of microswimming: instead of studying a complex, shape-deforming swimmer and the associated time-varying flow field, we can consider the motion of a constant-strength force dipole in the same environment. With this approximation, one can often derive surrogate equations of motion for the swimmer \cite{Kim2005,Lauga2009}, which can then be analysed with ease, at least when compared to models that capture the intricate, time-dependent details of the microswimmer and the flow that it generates.

Though many of the assumptions associated with this modelling approach are well understood, such as the limitations of the far-field approximation when studying near-field interactions, the impact of assuming a constant dipole strength is less clear. More generally, the impact of adopting constant, \textit{a priori}-averaged parameters in simple models of temporally evolving microswimmers has not been thoroughly investigated, to the best of our knowledge. However, it is clear that employing rapidly varying parameters can have significant consequences for the predictions of simple models. For example, the work of \citet{Omori2022} recently explored a simple model of shape-changing swimmers, explicitly including rapid variation in the parameters that describe the swimmer shape and its speed of self propulsion. The subsequent analysis of \citet{Walker2022a} highlighted, amongst other observations, that the fast variation in the parameters was key to the behavioural predictions of the model, which were found to align with the experimental observations of \citet{Omori2022}. Hence, the study of the effects of employing time-dependent parameters in even simple models, in comparison to adopting constant, averaged parameters, is warranted.

Thus, the primary aim of this study will be to explore minimal models of microswimming, incorporating fast variation in model parameters. To do this, motivated by a number of recent works in a similar vein \cite{Walker2022,Walker2022a,Gaffney2022}, we will employ a multiple scales asymptotic analysis \cite{Bender1999} to systematically derive effective governing equations from non-autonomous models. In particular, we will incorporate the effects of rapid variation by exploiting the separation of timescales often associated with microswimming, yielding leading-order autonomous dynamical systems. In what follows, we will focus on three example scenarios: the interaction of a dipole swimmer with a boundary, the hydrodynamic interaction of two mirror-symmetric dipole swimmers, and the angular dynamics of two hydrodynamically interacting dipoles that are pinned in place. Through our analysis, which will be simple, if not elementary, we will compare the dynamics of multi-timescale models with the predictions of the simplest, constant-parameter models, seeking to ascertain both if qualitative differences arise and if they can be systematically corrected for by informed parameter choices.

\section{A dipole near a no-slip boundary}\label{sec: no slip}
\subsection{Model equations}
\begin{figure}
    \centering
    \includegraphics[width=0.3\textwidth]{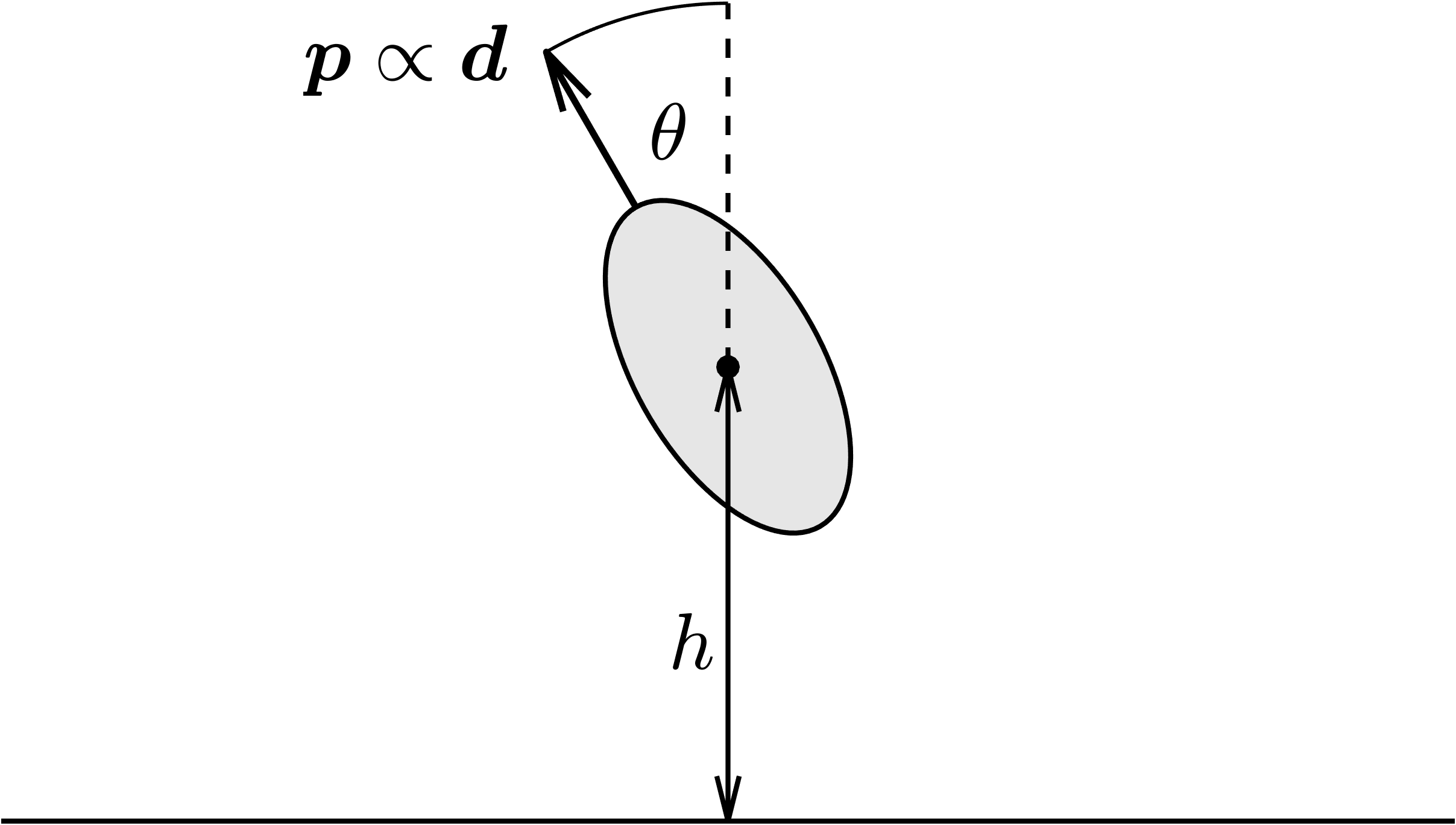}
    \caption{Geometry and parameterisation of a minimal model of a swimmer above a no-slip boundary. The swimmer is parameterised by the angle $\theta$ between the vector dipole strength $\p$ and the separation $h$ of its centroid from the boundary. The velocity due to self-propulsion is assumed to be along the direction of $\p$, shared with swimmer-fixed director $\d$.}
    \label{fig: no slip: setup}
\end{figure}

Consider a swimmer moving in a half space that is bounded by an infinite no-slip plane, with the swimmer moving in a plane perpendicular to the boundary. We parameterise the orientation of the swimmer by the angle $\theta$ between a swimmer-fixed director $\d$ and the boundary normal, and parameterise its position by the distance $h$ from its centre to the boundary, as illustrated in \cref{fig: no slip: setup}. With all quantities dimensionless, a minimal model for the swimmer dynamics is presented in part by \citet{Lauga2020} as
\begin{subequations}\label{eq: no slip: original system}
\begin{align}
    \diff{h}{t} &= \frac{3p}{16h^2}\left(1 + 3\cos{2\theta}\right) + u \cos{\theta}\,,\\
    \diff{\theta}{t} &= \frac{3p\sin{2\theta}}{64h^3}\left[4 + B(3 + \cos{2\theta})\right]\,,\label{eq: no slip: original system: angular}
\end{align}
\end{subequations}
where $u$ is the speed of self propulsion and we have shifted \citeauthor{Lauga2020}'s definition of $\theta$ by $\pi/2$. In this minimal model, the flow generated by the swimmer in the absence of the boundary is assumed to be purely that of a force dipole with vector strength $\p=p\d$, aligned along the body fixed director that defines $\theta$, and the swimmer shape is captured only through the Bretherton parameter $B$ \cite{Bretherton1962}. This modelling approach can be justified by considering a far-field limit of a swimmer, though here we focus on analysing the model of \cref{eq: no slip: original system} rather than on its origin and motivation. In particular, we focus on the angular dynamics contained within \cref{eq: no slip: original system: angular}.

The standard approach to modelling this system would be to assume that $p$, $u$, and $B$ are constant in time, as is the case in the textbook of \citet{Lauga2020}. This can be interpreted as averaging away any time dependence of the three parameters, which one would generically expect to be present for a multitude of shape-changing microswimmers, for instance. Here, we do not perform this \emph{\textit{a priori}} averaging of the parameters, and will instead suppose that $p$, $u$, and $B$ are indeed functions of time. In particular, we suppose that $p=p(\omega t)$, $u=u(\omega t)$, and $B=B(\omega t)$ are periodic functions of $\omega t$, where $\omega\gg1$ is a large dimensionless frequency of oscillation and we assume that $p$, $u$, and $B$ share a period, in line with the rapid shape changes undergone by many microswimmers. For later convenience, we make the additional assumption that the average of $p$ over a period is non-zero, and will impose the minimal restriction that $B\in(-1,1)$, which holds for all but the most elongated of objects \cite{Bretherton1962}. Hence, we study the non-autonomous system
\begin{subequations}\label{eq: no slip: full system}
\begin{align}
    \diff{h}{t} &= \frac{3p(\omega t)}{16h^2}\left(1 + 3\cos{2\theta}\right) + u(\omega t) \cos{\theta}\,,\\
    \diff{\theta}{t} &= \frac{3p(\omega t)\sin{2\theta}}{64h^3}\left[4 + B(\omega t)(3 + \cos{2\theta})\right]\,,
\end{align}
\end{subequations}
with $\omega\gg1$ and all other quantities being $\bigO{1}$ as $\omega\to\infty$. 

\subsection{Multi-scale analysis}\label{sec: no slip: multiscale analysis}
We will attempt to exploit the large frequency $\omega \gg 1$ in order to make progress in analysing the non-autonomous system of \cref{eq: no slip: full system}, employing the method of multiple scales \cite{Bender1999}. Following this approach, we introduce the fast timescale $T \coloneqq \omega t$, so that $p = p(T)$ etc., and formally treat $t$ and $T$ as independent. The proper time derivative $\mathrm{d}/\mathrm{d}t$ accordingly transforms as
\begin{equation}\label{eq: no slip: time transform}
    \diff{}{t} \mapsto \pdiff{}{t} + \omega \pdiff{}{T}\,,
\end{equation}
transforming our non-autonomous system of ordinary differential equations (ODEs) into a system of partial differential equations (PDEs). We now seek asymptotic expansions of $h$ and $\theta$ in inverse powers of $\omega$, which we write as
\begin{equation}
    h \sim h_0(t,T) + \frac{1}{\omega}h_1(t,T) + \cdots\,, \quad \theta \sim \theta_0(t,T) + \frac{1}{\omega}\theta_1(t,T) + \cdots\,.
\end{equation}
Transforming \cref{eq: no slip: full system} via \cref{eq: no slip: time transform} and inserting these asymptotic expansions gives the $\bigO{\omega}$ balance simply as
\begin{equation}
    \pdiff{h_0}{T} = 0\,, \quad \pdiff{\theta_0}{T} = 0 \quad \implies \quad h_0 = h_0(t)\,, \quad \theta_0 = \theta_0(t)\,,
\end{equation}
so that the leading-order solutions are independent of the fast timescale $T$. This should be expected, as the forcing of the system is strictly $\bigO{1}$, so that the dominant contribution to the evolution occurs on the long timescale $t$.

At the next asymptotic order, we pick up the $\bigO{1}$ forcing and we have
\begin{subequations}\label{eq: no slip: order unity system}
\begin{align}
    \diff{h_0}{t} + \pdiff{h_1}{T} &= \frac{3p(T)}{16h_0^2}(1 + 3\cos{2\theta_0}) + u(T)\cos{\theta_0}\,,\\
    \diff{\theta_0}{t} + \pdiff{\theta_1}{T} &= \frac{3p(T)\sin{2\theta_0}}{64h_0^3}\left[4 + B(T)(3 + \cos{2\theta_0})\right]\,,
\end{align}
\end{subequations}
writing $t$-derivatives of $h_0$ and $\theta_0$ as proper due to their established independence from $T$. The appropriate solvability conditions for this first-order system are obtained by averaging the equations over a period in $T$ and imposing periodicity in $T$, equivalent to the Fredholm Alternative Theorem for this system \cite{Bender1999}. To do so, we assume, without loss of generality, that the period of the fast oscillations is $2\pi$, defining the averaging operator $\avg{\cdot}$ via
\begin{equation}\label{eq: no slip: averaging operator}
    \avg{a} = \frac{1}{2\pi}\int_0^{2\pi}a(T)\intd{T}\,.
\end{equation}
Computing the average of \cref{eq: no slip: order unity system}, we arrive at
\begin{subequations}\label{eq: no slip: systematic model}
\begin{align}
    \diff{h_0}{t} &= \frac{3\avg{p}}{16h_0^2}(1 + 3\cos{2\theta_0}) + \avg{u}\cos{\theta_0}\,,\\
    \diff{\theta_0}{t} &= \frac{3\avg{p}\sin{2\theta_0}}{64h_0^3}\left[4 + \frac{\avg{pB}}{\avg{p}}(3 + \cos{2\theta_0})\right]\,,
\end{align}
\end{subequations}
with the imposed periodicity eliminating the fast-time derivatives. Comparing these leading-order differential equations with those of \cref{eq: no slip: full system}, we see that we have essentially replaced the parameters $p$, $u$, and $B$ with the effective parameters $\avg{p}$, $\avg{u}$, and $\avg{pB}/\avg{p}$, the precise forms of which have arisen through our brief, systematic analysis. Whilst the modifications to $p$ and $u$ are as might be naively expected, the effective shape constant $\avg{pB}/\avg{p}$ is somewhat less obvious at first glance, with one perhaps expecting the average parameter $\avg{B}$. Indeed, these authors have previously been guilty of utilising simply $\avg{p}$, $\avg{u}$, and $\avg{B}$ in place of the rapidly oscillating quantities in back-of-the-envelope calculations. We will refer to such a model as an \emph{\textit{a priori}-averaged model}, here given explicitly by 
\begin{subequations}\label{eq: no slip: a priori model}
\begin{align}
    \diff{h^a}{t} &= \frac{3\avg{p}}{16{h^a}^2}\left(1 + 3\cos{2\theta^a}\right) + \avg{u} \cos{\theta^a}\,,\\
    \diff{\theta^a}{t} &= \frac{3\avg{p}\sin{2\theta^a}}{64{h^a}^3}\left[4 + \avg{B}(3 + \cos{2\theta^a})\right]\,,
\end{align}
\end{subequations}
using a superscript of $a$ to denote the solutions of this \textit{a priori}-averaged system.

Though an elementary observation, it is worth highlighting that, without any additional assumptions on $p(T)$ and $B(T)$, it is in general not the case that $\avg{pB}/\avg{p} = \avg{B}$, so that we should expect to observe differences between the systematically determined, leading-order dynamics of \cref{eq: no slip: systematic model} and those of the \textit{a priori}-averaged model of \cref{eq: no slip: a priori model}. In what follows, through a brief consideration of the angular evolution equations, we will highlight how these differences can be more than simply quantitative.

Focussing on the angular dynamics, we specifically consider the abstracted scalar autonomous ODE
\begin{equation}\label{eq: no slip: scalar ode}
    \diff{x}{t} = f(x;\alpha,\beta)\,,
\end{equation}
where
\begin{equation}\label{eq: no slip: f}
    f(x;\alpha,\beta) \coloneqq \frac{3\alpha\sin{2x}}{64h^3}\left[4 + \beta(3 + \cos{2x})\right]\,,
\end{equation}
so that $(\alpha,\beta) = (\avg{p},\avg{B})$ corresponds to the \textit{a priori}-averaged model, whilst $(\alpha,\beta) = (\avg{p},\avg{pB}/\avg{p})$ gives the leading-order, systematically averaged dynamics. Note that, for the purposes of a stability analysis of the angular dynamics, we can treat the swimmer separation from the boundary as a positive parameter, abusing notation and generically writing $h$ in the denominator of \cref{eq: no slip: f}, without materially modifying a steady state analysis of the angular dynamics.

\begin{figure}
    \centering
    \begin{overpic}[permil,width=0.7\textwidth]{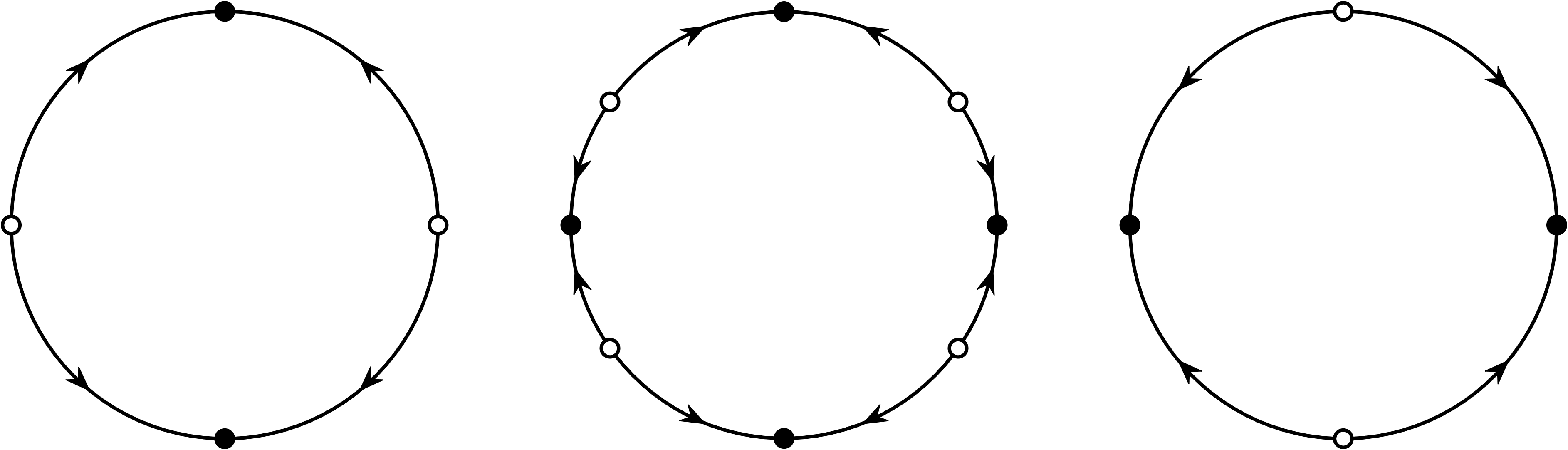}
    \put(100,140){$\beta<-2$}
    \put(425,140){$-2<\beta<-1$}
    \put(810,140){$\beta>-1$}
    \put(0,300){(a)}
    \put(350,300){(b)}
    \put(700,300){(c)}
    \put(115,300){$x=0$}
    \put(-30,140){$\frac{\pi}{2}$}
    \put(136,-27){$\pi$}
    \put(293,140){$\frac{3\pi}{2}$}
    \end{overpic}
    \caption{Steady states and stability of angular evolution for the autonomous system of \cref{eq: no slip: scalar ode} are shown as dynamics on a circle, for $\alpha>0$ fixed and for three values of $\beta$. Swimming parallel to the boundary corresponds to states with $x=\pi/2$, $3\pi/2$, whilst $x=0$, $\pi$ corresponds to swimming aligned with the normal to the boundary. (a) With $\beta < -2$, the system evolves to a steady state with $x=n\pi$, $n\in\mathbb{Z}$. (b) For $-2<\beta<-1$, $x=n\pi/2$ are stable for all $n\in\mathbb{Z}$, with unstable configurations present between these attractors. (c) For $\beta > -1$, the system evolves to a steady state with $x=\pi/2 + n\pi$, for $n\in\mathbb{Z}$. Stable states are shown as solid points, whilst unstable points are shown hollow. Stabilities for $\alpha<0$ are obtained by reversing the illustrated dynamics.}
    \label{fig: no slip: dynamics}
\end{figure}
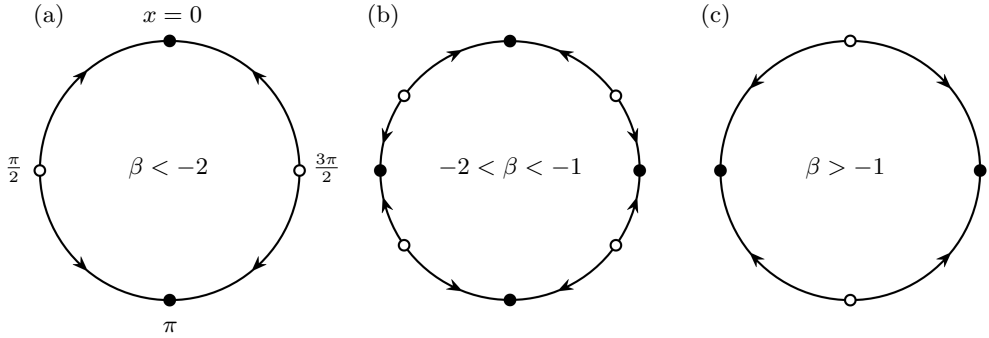

\subsection{Exploring the autonomous dynamics}
The fixed points of \cref{eq: no slip: scalar ode} are readily seen to be $x = n\pi$, $x=\pi/2 + n\pi$, and solutions of $4 + \beta(3 + \cos{2x})=0$ (if they exist), for $n\in\mathbb{Z}$. Notably, if $\beta\in(-1,1)$, as is the case in the \textit{a priori}-averaged model, then the only steady states are at integer multiples of $\pi/2$. Focussing on these steady states, their linear stability is given by
\begin{equation}\label{eq: no slip: simple steady states}
    x = \left\{\begin{array}{ll}
        n\pi & \text{ is stable } \iff\alpha(1+\beta) < 0\,,\\
        \pi/2 + n\pi & \text{ is stable } \iff\alpha(2+\beta) > 0\,.\\
        \end{array}\right.
\end{equation}
Hence, for $\beta\in(-1,1)$, the stability of the steady states is determined solely by the sign of $\alpha$, with $\alpha>0$ giving rise to unstable states at $x=n\pi$ and stable states at $x=\pi/2+n\pi$. Identifying $\alpha$ with the averaged signed dipole strength, this is precisely in line with the classical analysis of pusher and puller swimmers via the \textit{a priori}-averaged model, as summarised by \citet{Lauga2009}, with pushers and pullers corresponding to $\alpha > 0$ and $\alpha < 0$, respectively.

However, if $\beta<-1$, the profile of stability can change significantly. Bifurcations at $\beta=-1$ and $\beta=-2$ see the creation and destruction of additional steady states (the solutions of $4 + \beta(3 + \cos{2x})=0$) accompanied by changes in stability of the steady states at $x=n\pi$ and $x=\pi/2 + n\pi$, respectively. When they exist, the additional states have the opposite stability to the other steady states, so that they are stable for $\alpha<0$. For $\beta<-2$, the equation defining the additional steady states admits no real solutions, so that these steady states cease to exist and the angular equilibria are the same as for $\beta>-1$, though with opposite linear stabilities. Each of these dynamical regimes is illustrated in \cref{fig: no slip: dynamics} for $\alpha>0$, highlighting a strong dependence of the dynamics on $\beta$. The linear stability of each state is flipped upon taking $\alpha<0$.

\subsection{Comparing the emergent dynamics}
The elementary analysis of the previous section highlights how qualitative changes in the globally attractive behaviour of the model depend strongly on the parameter $\beta$. However, the predictions of the \textit{a priori}-averaged model are simple: if the swimmer is a pusher, with $\alpha=\avg{p}>0$, then the states $\theta^a=n\pi$ are unstable, and the swimmer instead evolves to a state where $\theta^a = \pi/2 + n\pi$ and swims parallel to the boundary. If the swimmer is a puller, with $\alpha=\avg{p}<0$, then the swimmer instead evolves to $\theta^a=n\pi$, thereafter moving perpendicular to the boundary.

However, the predictions of the systematically averaged model are more complex. Whilst switching the sign of $\avg{p}$ still switches the stability of each steady state, the value of $\beta = \avg{pB}/\avg{p}$, which need not be smaller than unity in magnitude, can materially alter both the steady states in existence and the stability of the states that correspond to parallel and perpendicular swimming. For instance, if $\alpha=\avg{p}>0$ and $\beta\in(-2,-1)$, both the $\theta_0=n\pi$ and the $\theta_0=\pi/2+n\pi$ states are linearly stable, accompanied by four steady states in the range $\theta_0\in(0,2\pi)$ that are unstable; for $\alpha=\avg{p}<0$, the stability of each state is swapped. Hence, swimmers with $\avg{p}<0$ and $\avg{pB}/\avg{p}\in(-2,-1)$ will evolve to a steady state that is not a multiple of $\pi/2$; in other words, they will neither align parallel nor perpendicular to the boundary, a behaviour that is never predicted by the \textit{a priori}-averaged model in any admissible parameter regime.

\begin{figure}
    \centering
    \includegraphics[width=0.3\textwidth]{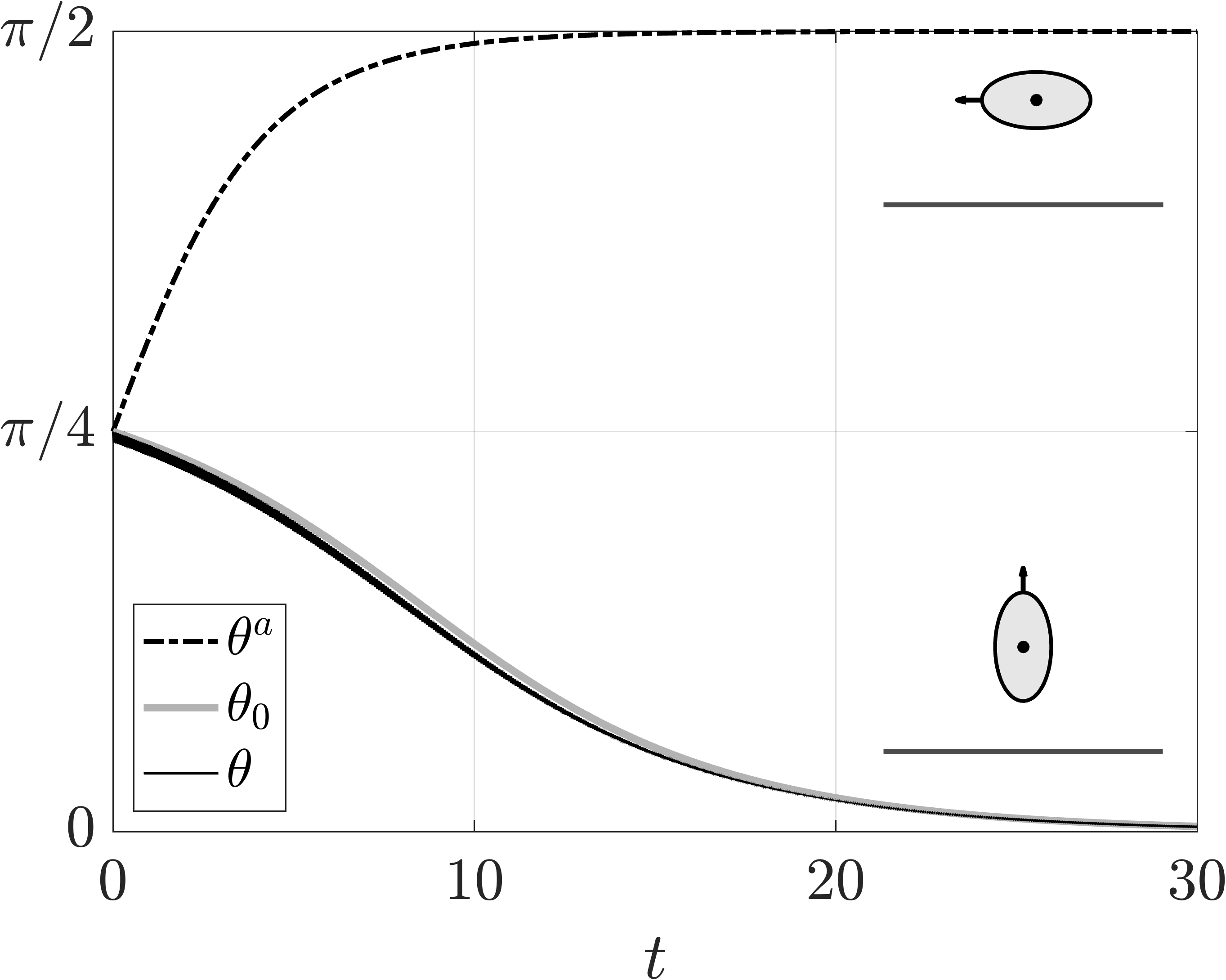}
    \caption{Angular evolution of a swimmer at fixed separation from a no-slip boundary. The prediction of the \textit{a priori}-averaged model can be seen to not align with the prediction of the systematically averaged model of \cref{eq: no slip: systematic model} or the dynamics of the full model of \cref{eq: no slip: full system}, with qualitatively distinct steady configurations from the same initial condition. Rapid oscillations in the solution full dynamics are not visible at the resolution of this plot. Here, we have fixed $h=1$, $\omega = 100$, and taken $p(T) = -6\sin{(T)} + 1$ and $B(T) = \sin{(T)}/2$. Schematics of the long-time configurations are shown inset.}
    \label{fig: no slip: example}
\end{figure}

As an explicit illustration of how the two models can qualitatively differ, we take $p(T) = 4A\sin{T} + 1$ and $B(T) = \sin{(T)}/2$, so that $\avg{p} = 1$, $\avg{B}=0$, and $\avg{pB}/\avg{p} = A$. The \textit{a priori}-averaged model predicts the dynamics shown in \cref{fig: no slip: dynamics}c for all values of $A$, whilst the systematically averaged dynamics follow \cref{fig: no slip: dynamics}b for $A\in(-2,-1)$ and \cref{fig: no slip: dynamics}a for $A<-2$. Fixing $A=-3/2$, the temporal evolution of both of these models is shown in \cref{fig: no slip: example}, along with a numerical solution to the angular dynamics of the full system of \cref{eq: no slip: full system}. Here, we have fixed $h>0$ as a parameter, recalling that the swimmer separation serves only to modify the rate at which the angular dynamics approach a steady state. In agreement with our analysis, the \textit{a priori}-averaged model incorrectly predicts the qualitative evolution of the model swimmer, whilst the leading-order, systematically averaged model is in agreement with the full numerical solution.

Despite these general qualitative differences, it should be noted that there are parameter regimes in which the dynamics are qualitatively indistinct between the models. For instance, suppose that we are in a regime where $\beta=\avg{pB}/\avg{p}\not\in(-2,-1)$, so that the only steady states are those given in \cref{eq: no slip: simple steady states}. Then, if $p(T)$ is of fixed sign for all $T$, so that the swimmer is unambiguously a pusher or a puller, it is simple to note that $\avg{p} + \avg{pB} =\int_0^{2\pi}p(T)[1+B(T)]\intd{T}/2\pi$ has the same sign as $\avg{p}$, recalling that $B\in(-1,1)$. Similarly, $2\avg{p} + \avg{pB}$ has the same sign as $\avg{p}$, so that the long-time behaviour is completely determined by the sign of $\avg{p}$, noting the conditions from \cref{eq: no slip: simple steady states}.

\section{Mirror symmetric dipoles and a free-slip boundary}\label{sec: free slip}
\subsection{Model equations}\label{sec: free slip: model eqs}
\begin{figure}
    \centering
    \includegraphics[width=0.25\textwidth]{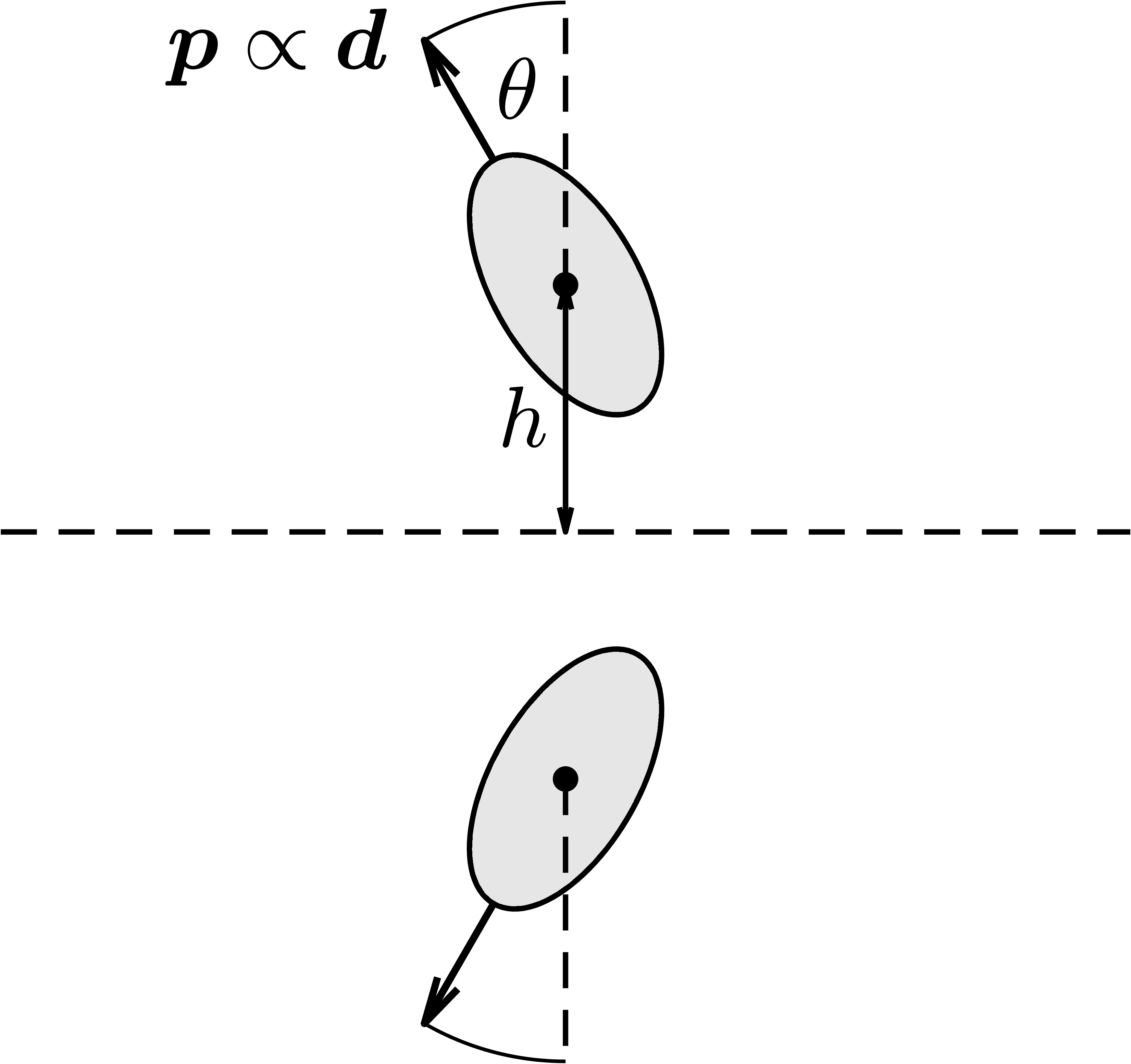}
    \caption{Geometry and parameterisation of interacting dipole swimmers undergoing mirror-symmetric motion, equivalent to the motion of one dipole in the presence of a free-slip boundary. The orientation of the swimmers is described by an angle $\theta$ between the vector dipole strength $\p$ and the boundary normal, with the separation between each swimmer and the plane of mirror symmetry being denoted by $h$. As in the previous section, the swimming velocity of each particle is assumed to align with the vector dipole strength.}
    \label{fig: free slip: setup}
\end{figure}

Consider the parameterised dipole swimmer of the previous section, with the same set-up and parameterisation as before, and modify the no-slip boundary condition to be that of a free-slip interface. This no-shear-stress boundary condition is well-known to be equivalent to imposing a symmetry condition on the flow and geometry, so that we can instead cast this problem in the context of two dipoles moving in a shared domain in a mirror-symmetric way. In this scenario, as illustrated in \cref{fig: free slip: setup}, the evolution of $h$ and $\theta$ is governed by the ODE system
\begin{subequations}\label{eq: free slip: original system}
\begin{align}
    \diff{h}{t} &= \frac{p}{8h^2}(1 + 3\cos{2\theta}) + u \cos{\theta}\,,\\
    \diff{\theta}{t} &= \frac{3p\sin{2\theta}}{32h^3}\left[2 + B(1 + \cos{2\theta})\right]\,,\label{eq: free slip: original system: angular}
\end{align}
\end{subequations}
following the description of \citet{Spagnolie2012} and qualitatively resembling the governing equations of \cref{sec: no slip}. Here, $p$, $u$, and $B$ are once again the signed dipole strength, the signed swimming speed, and the Bretherton parameter associated with the swimmer.

As with the previous model, a standard approach would be to characterise the dynamics of a dipole swimmer by assuming that the three quantities $p$, $u$, and $B$ were constant in time. Here, we will consider the case where these parameters oscillate rapidly in time, studying the non-autonomous dynamical system
\begin{subequations}\label{eq: free slip: non-autonomous system}
\begin{align}
    \diff{h}{t} &= \frac{p(\omega t)}{8h^2}(1 + 3\cos{2\theta}) + u(\omega t) \cos{\theta}\,,\\
    \diff{\theta}{t} &= \frac{3p(\omega t)\sin{2\theta}}{32h^3}\left[2 + B(\omega t)(1 + \cos{2\theta})\right]\,,\label{eq: free slip: non-autonomous system: angular}
\end{align}
\end{subequations}
for $\omega\gg1$ and all other quantities $\bigO{1}$ as $\omega\to\infty$. We will continue to assume that $B\in(-1,1)$ and that the average of $p$ is non-zero.

\subsection{Multi-scale analysis}
As \cref{eq: free slip: non-autonomous system} differs from the non-autonomous system of \cref{eq: no slip: full system} only through the simple modification of constants, the structure of the problem and a subsequent multi-scale analysis is entirely similar to the previous case. Following the approach taken in \cref{sec: no slip: multiscale analysis} with only trivial alterations leads to the systematically derived, leading-order system
\begin{subequations}\label{eq: free slip: systematic model}
\begin{align}
    \diff{h_0}{t} &= \frac{\avg{p}}{8h_0^2}(1 + 3\cos{2\theta_0}) + \avg{u}\cos{\theta_0}\,,\\
    \diff{\theta_0}{t} &= \frac{3\avg{p}\sin{2\theta_0}}{32h_0^3}\left[2 + \frac{\avg{pB}}{\avg{p}}(1 + \cos{2\theta_0})\right]\,,
\end{align}
\end{subequations}
where $h_0$ and $\theta_0$ are the leading-order terms in expansions of $h$ and $\theta$, respectively, and $\avg{\cdot}$ denotes averages taken over the shared period of fast oscillation of $p$, $u$, and $B$. Again, we see that this resembles the original dynamical system, with parameters replaced by their appropriately averaged counterparts. As before, though perhaps now less surprising, the Bretherton parameter $B$ has been replaced by $\avg{pB}/\avg{p}$, which need not align with $B$ in general. In order to explore the effects that this change of parameterisation can have on the predicted dynamics, we continue to interpret a model with constant parameters as an \textit{a priori}-averaged model of a temporally varying dipole swimmer, which can be stated explicitly as
\begin{subequations}\label{eq: free slip: naive system}
\begin{align}
    \diff{h^a}{t} &= \frac{\avg{p}}{8{h^a}^2}(1 + 3\cos{2\theta^a}) + \avg{u} \cos{\theta^a}\,,\\
    \diff{\theta^a}{t} &= \frac{3\avg{p}\sin{2\theta^a}}{32{h^a}^3}\left[2 + \avg{B}(1 + \cos{2\theta^a})\right]\,,\label{eq: free slip: naive system: angular}
\end{align}
\end{subequations}
adopting pre-averaged parameters in \cref{eq: free slip: original system} and denoting the solutions by $h^a$ and $\theta^a$.

Noting that the separation from the boundary acts only as a timescale in the angular evolution equations of both the systematically averaged dynamics and the \textit{a priori}-averaged model, we treat the separation $h$ as a parameter and consider only the angular dynamics, defining
\begin{equation}\label{eq: free slip: f}
    f(x;\alpha,\beta) \coloneqq \frac{3\alpha\sin{2x}}{32h^3}\left[2 + \beta(1 + \cos{2x})\right]
\end{equation}
and the autonomous scalar ODE
\begin{equation}\label{eq: free slip: scalar ode}
    \diff{x}{t} = f(x;\alpha,\beta)\,.
\end{equation}
In this setting, the \textit{a priori}-averaged model corresponds to taking $(\alpha,\beta) = (\avg{p},\avg{B})$, whilst the systematically averaged dynamics correspond to $(\alpha,\beta) = (\avg{p},\avg{pB}/\avg{p})$.

\subsection{Exploring the autonomous dynamics}
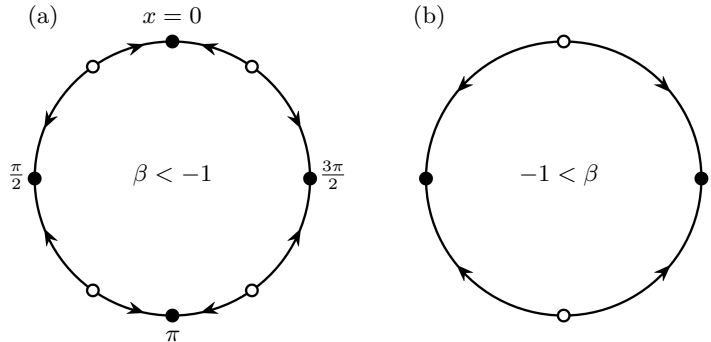
\begin{figure}
    \centering
    \begin{overpic}[permil,width=0.5\textwidth]{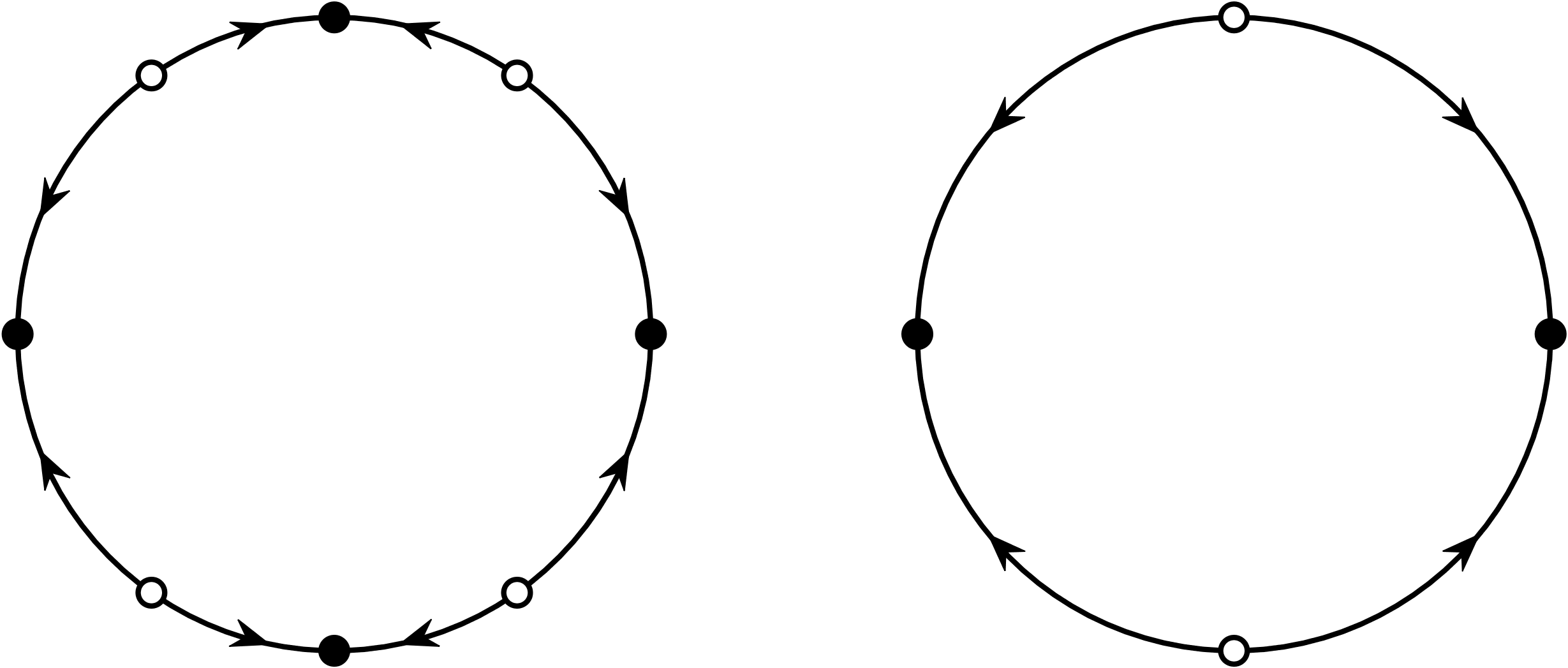}
    \put(155,210){$\beta < -1$}
    \put(722,210){$-1 < \beta$}
    \put(0,443){(a)}
    \put(566,443){(b)}
    \put(169,443){$x=0$}
    \put(-30,210){$\frac{\pi}{2}$}
    \put(203,-27){$\pi$}
    \put(428,210){$\frac{3\pi}{2}$}
    \end{overpic}
    \caption{Steady states and stability of angular evolution for the autonomous system of \cref{eq: free slip: scalar ode} are shown as dynamics on a circle, for $\alpha>0$ fixed and for various values of $\beta$. Swimming parallel to one another corresponds to states with $x=\pi/2$, $3\pi/2$, whilst $x=0$, $\pi$ corresponds to swimming towards or away from the other swimmer. (a) For $\beta < -1$, $x=n\pi/2$ are stable for all $n\in\mathbb{Z}$, with unstable configurations present between these attractors. (b) For $\beta > -1$, the system evolves to a steady state with $x=\pi/2 + n\pi$, for $n\in\mathbb{Z}$. Stable states are shown as solid points, whilst unstable points are shown hollow. Stabilities for $\alpha<0$ are obtained by reversing the illustrated dynamics.}
    \label{fig: free slip: dynamics}
\end{figure}
The autonomous dynamics of \cref{eq: free slip: scalar ode} closely resemble those of the no-slip case. The fixed points of the dynamics are given by $x = n\pi$, $x=\pi/2 + n\pi$, and solutions of $2 + \beta(1 + \cos{2x})=0$, for $n\in\mathbb{Z}$ and solutions to the latter relation existing precisely when $\beta\leq-1.$ Hence, for the \textit{a priori}-averaged model, where $\beta\in(-1,1)$, the only steady states are integer multiples of $\pi/2$, as was the case for the no-slip dynamics. The linear stability of these states is given by
\begin{equation}\label{eq: free slip: simple steady states}
    x = \left\{\begin{array}{ll}
        n\pi & \text{ is stable } \iff\alpha(1+\beta) < 0\,,\\
        \pi/2 + n\pi & \text{ is stable } \iff\alpha > 0\,.\\
        \end{array}\right.
\end{equation}
If $\beta\in(-1,1)$, we again see that the sign of $\alpha$ determines the stable configuration, with $\alpha = \avg{p} > 0$ corresponding to the prediction that classically described pusher swimmers will swim alongside one another, with $x=\pi/2 + n\pi$ being the only steady states of the angular dynamics. Analogously, puller swimmers are predicted to stably align along the axis of their dipoles, swimming directly away from or towards one another.

However, if $\beta < -1$, the stability of the $x=n\pi$ steady states switches. For $\alpha<0$, these states become unstable as $\beta$ crosses $-1$ from above, with their stability lost to the now-real solutions of $2 + \beta(1 + \cos{2x})=0$. The stability of the parallel-swimming states, with $x=\pi/2 + n\pi$, is unaffected by $\beta$, so that $\alpha >0$, $\beta < -1$ corresponds to a regime in which both $x=n\pi$ and $x=\pi/2 + n\pi$ are linearly stable, separated by the unstable solutions to $2 + \beta(1 + \cos{2x})=0$. Unlike in the analysis of the no-slip problem in the previous section, this regime in which all possible steady states coexist persists for all $\beta < -1$, so that the full range of dynamics can be captured using only two portraits, as shown in \cref{fig: free slip: dynamics} for the case with $\alpha>0$. As before, the stability of all steady states is switched upon changing the sign of $\alpha$, whilst changes in $\beta$ enact qualitative changes in the profile of stability.

\subsection{Comparing the emergent dynamics}
The above analysis highlights how the parameter $\beta$ again plays a key role in the dynamics, with changes in $\beta$ able to qualitatively alter the swimming behaviour. As in the previous case, the \textit{a priori}-averaged model predicts simply that pusher swimmers, with $\alpha=\avg{p}>0$, will swim stably with $\theta^a = \pi/2 + n\pi$, whilst puller swimmers swim with $\theta^a = n\pi$, so that pullers are predicted to align perpendicular to free-slip boundaries.

\begin{figure}
    \centering
    \includegraphics[width=0.3\textwidth]{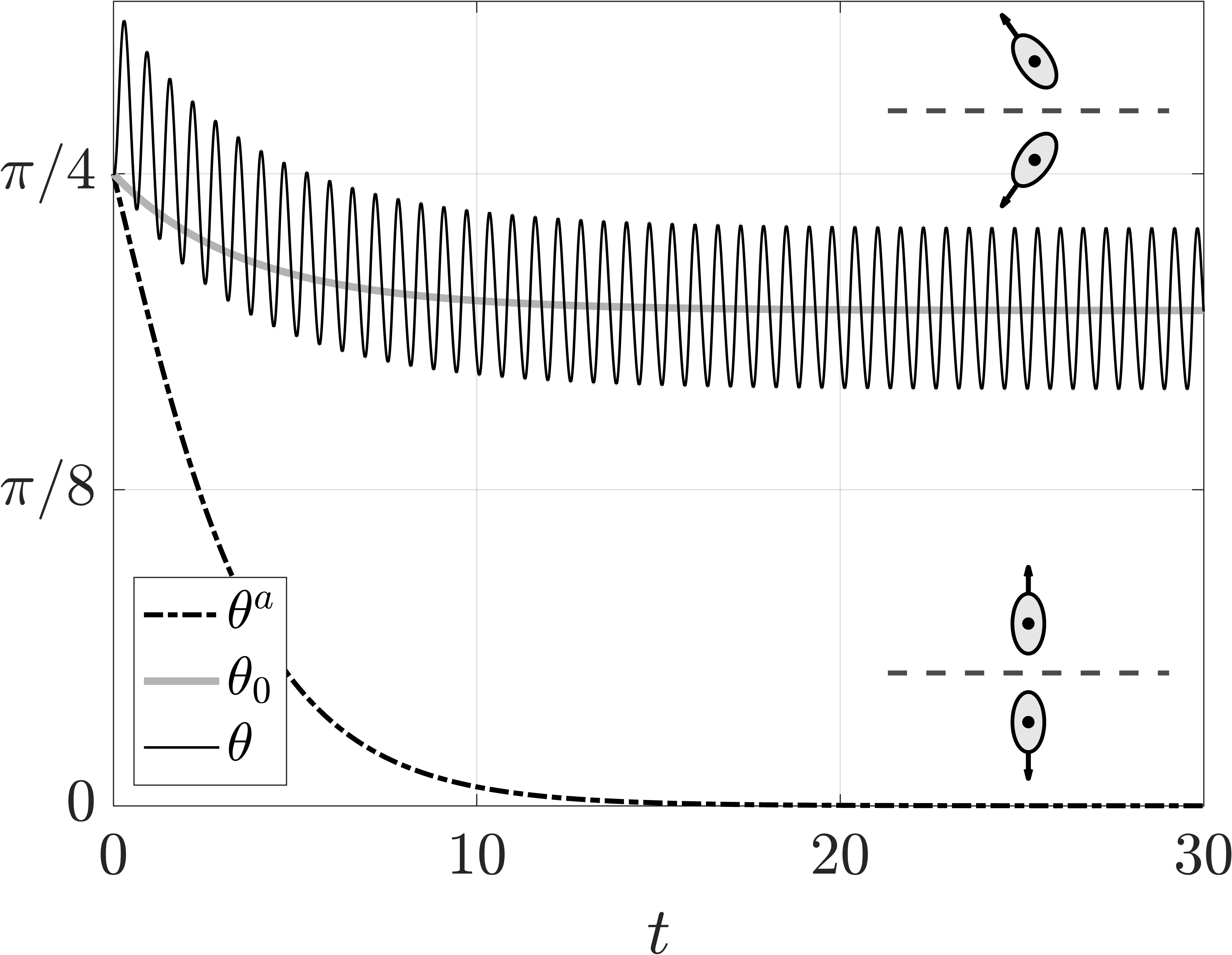}
    \caption{Angular evolution of a swimmer at fixed separation from a free-slip boundary. The prediction of the \textit{a priori}-averaged model can be seen to not align with the prediction of the systematically averaged model of \cref{eq: free slip: systematic model} or the dynamics of the full model of \cref{eq: free slip: non-autonomous system}, with qualitatively distinct steady configurations from the same initial condition. Here, we have fixed $h=1$, $\omega = 10$, and taken $p(T) = 6\sin{(T)} - 1$ and $B(T) = \sin{(T)}/2$. Schematics of the long-time configurations are shown inset.}
    \label{fig: free slip: example}
\end{figure}

The predictions of the systematically averaged model are more complex, with $\beta=\avg{pB}/\avg{p}$ no longer constrained to be of magnitude less than unity. For example, in the case where $\beta<-1$, \cref{fig: free slip: dynamics}a highlights that pusher swimmers, with $\alpha=\avg{p}>0$, can evolve any configuration with $\theta^a = n\pi/2$, so that they can in fact be stable both parallel and perpendicular to a free-slip boundary. This prediction is clearly at odds with that of the \textit{a priori}-averged model, evidencing the potential unreliability of the simplest microswimmer models when inappropriately parameterised. Puller swimmers, on the other hand, exhibit somewhat more interesting behaviour for $\beta<-1$, evolving to a state neither perpendicular nor parallel to the boundary, in contrast to the perpendicular configuration predicted by the \textit{a priori}-averaged model. This example is illustrated numerically in \cref{fig: free slip: example}.

As before, there are classes of $p(T)$ and $B(T)$ for which the qualitative predictions of both models align. Specifically, if $p(T)$ is of fixed sign for all $T$, then it is the case that $\avg{p}$ and $\avg{p} + \avg{pB}$ have the same sign, so that the stability conditions of \cref{eq: free slip: simple steady states} depend only on $\alpha=\avg{p}$, which is the same in both models.

\section{Interacting dipole rollers}\label{sec: rollers}
\subsection{Model equations}\label{sec: rollers: model eqs}
Consider a pair of particles in the plane that are pinned in place in the laboratory frame, so that they are free to rotate in the plane but unable to translate. The particles, which we term \emph{rollers}, are assumed to interact through dipolar flow fields, with vector dipole strengths $\p_1$ and $\p_2$, respectively, which are both assumed to lie in the plane containing the particles. Taking $\{\ex,\ey\}$ to be an orthogonal basis for the laboratory frame that spans this plane, we define the orientation of the $i$\textsuperscript{th} particle, $i\in\{1,2\}$, via the angle $\theta_i$ between a body-fixed axis and $\ex$, so that the direction of the roller can be captured as $\d_i = \cos{\theta_i}\ex + \sin{\theta_i}\ey$. Analogously to the previous sections, we assume that the vector dipole strength of each particle is aligned with $\d_i$, so that we can write $\p_i = p_i\d_i$, where $p_i$ is the scalar dipolar strength and may take any sign. We further assume that $p_1=p_2$, so that the dipoles are of equal strength, though we remark that retaining generality is straightforward but notationally cumbersome.
\begin{figure}
    \centering
    \includegraphics[width=0.4\textwidth]{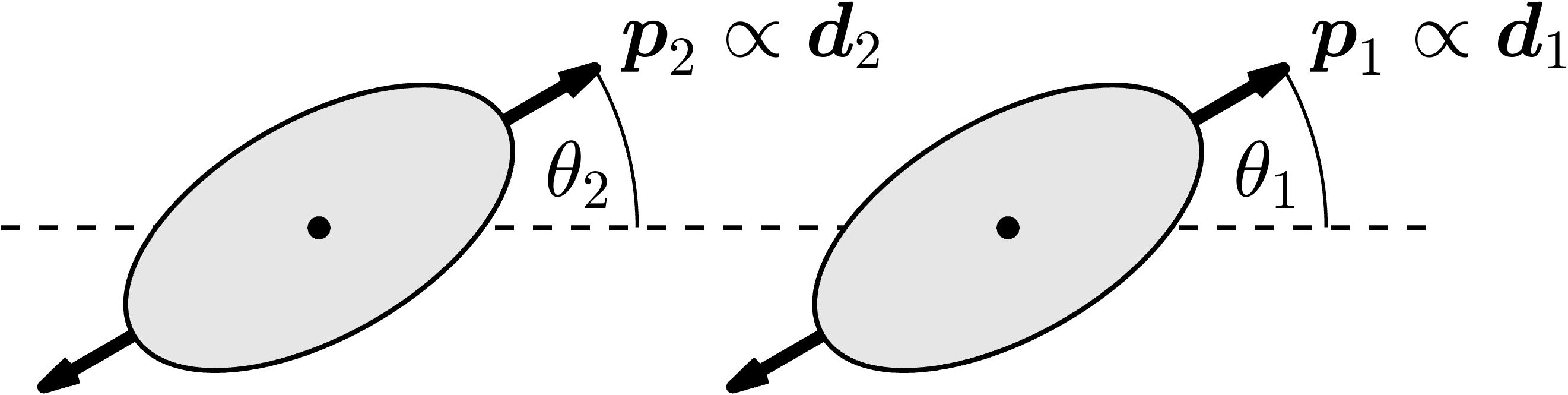}
    \caption{Geometry of fixed-position dipole rollers. Each roller is pinned in place at its centre, but is free to rotate, driven by the dipolar flow generated by the other. Their orientation is captured by their directors $\d_1$ and $\d_2$, parameterised by $\theta_1$ and $\theta_2$, respectively. The axis of each roller's dipolar flow field is assumed to be directed along their orientation vector.}
    \label{fig: roller setup}
\end{figure}
Without loss of generality, we assume that the displacement of particle $1$ from particle $2$ is $r\ex$, where $r>0$ is the distance between them. In this setting, illustrated in \cref{fig: roller setup} and following \citet[p. 295]{Lauga2020}, the effects of each particle's dipolar flow on the orientation of the other particle are captured by the following dimensionless coupled system of ODEs:
\begin{subequations}\label{eq: rollers: full system}
\begin{align}
    \diff{\theta_1}{t} &= f(\theta_1,\theta_2;p,B)\,,\\
    \diff{\theta_2}{t} &= f(\theta_2,\theta_1;p,B)\,,
\end{align}
\end{subequations}
where
\begin{equation}\label{eq: rollers: def f}
    f( x,y;\alpha,\beta) \coloneqq -\frac{3\alpha}{r^3}\left(\sin{ y}\cos{ y} + \beta\left[\sin{ x}\cos{ x}\left(1 - 5\cos^2{ y}\right) + \cos{ y}\sin{(2 x- y)}\right]\right)\,.
\end{equation}
Here, $B$ is once again the shape-capturing parameter of \citet{Bretherton1962}, and we assume that both of the particles are associated with the same shape parameter. This further assumption can be relaxed at the expense of notational convenience.

The standard approach would be to assume that $p$ and $B$ are constant, so that the above system is autonomous. Here, we take $p=p(T)$ and $B=B(T)$, interpreting the constant-parameter model as the \textit{a priori}-averaged model, as above. Hence, we study the non-autonomous system
\begin{subequations}\label{eq: rollers: non-autonomous system}
\begin{align}
    \diff{\theta_1}{t} &= f(\theta_1,\theta_2;p(\omega t),B(\omega t)) \,,\\
    \diff{\theta_2}{t} &= f(\theta_2,\theta_1;p(\omega t), B(\omega t))\,,
\end{align}
\end{subequations}
with $\omega\gg1$ and all other quantities being $\bigO{1}$ as $\omega\to\infty$. It should be noted that, whilst we study this system as a simple extension of an established model, these equations can be rigorously derived by considering the dynamics of shape-changing spheroids, subject to the far-field approximation and the assumption that their instantaneous deformation is both force- and torque-free. This derivation is somewhat elementary and follows the approach given in \citet{Gaffney2022b}, from which this model can also be extended to accommodate more general deformations through the addition of uncomplicated terms. Here, seeking simplicity and clarity, we pursue the model given in \cref{eq: rollers: non-autonomous system}.

\subsection{Multi-scale analysis}
The multi-scale analysis of this problem proceeds entirely analogously to that of the previous two examples, but is reproduced here in detail in the interest of clarity. As in \cref{sec: no slip}, we formally introduce the fast timescale $T=\omega t$, transforming the proper time derivative as
\begin{equation}\label{eq: rollers: time transform}
    \diff{}{t} \mapsto \pdiff{}{t} + \omega \pdiff{}{T}\,.
\end{equation}
We seek an asymptotic expansion of $\theta_1$ and $\theta_2$ in inverse powers of $\omega$, which we write as
\begin{equation}\label{eq: rollers: expansions}
    \theta_1 \sim \theta_{1,0}(t,T) + \frac{1}{\omega}\theta_{1,1}(t,T) + \cdots\,, \quad \theta_2 \sim \theta_{2,0}(t,T) + \frac{1}{\omega}\theta_{2,1}(t,T) + \cdots\,.
\end{equation}
Transforming \cref{eq: rollers: non-autonomous system} via \cref{eq: rollers: time transform} and inserting the expansions of \cref{eq: rollers: expansions}, at $\bigO{\omega}$ we have
\begin{equation}
    \pdiff{\theta_{1,0}}{T} = 0\,, \quad \pdiff{\theta_{2,0}}{T} = 0 \quad \implies \quad \theta_{1,0} = \theta_{1,0}(t)\,, \quad \theta_{2,0} = \theta_{2,0}(t)\,,
\end{equation}
so that the leading-order solutions for $\theta_1$ and $\theta_2$ are independent of the fast timescale and, thus, are functions only of $t$. As in \cref{sec: no slip,sec: free slip}, this arises due to the forcing of the ODEs being $\bigO{1}$, so that the forcing does not contribute at $\bigO{\omega}$. At $\bigO{1}$, we have
\begin{subequations}\label{eq: rollers: first-order system}
\begin{align}
    \diff{\theta_{1,0}}{t} + \pdiff{\theta_{1,1}}{T} &= f(\theta_{1,0},\theta_{2,0};p(T), B(T)) \,,\label{eq: rollers: first-order system: theta1}\\
    \diff{\theta_{2,0}}{t} + \pdiff{\theta_{2,1}}{T} &= f(\theta_{2,0},\theta_{1,0};p(T), B(T))\,,\label{eq: rollers: first-order system: theta2}
\end{align}
\end{subequations}
using the fact that $\theta_{1,0}$ and $\theta_{2,0}$ are independent of $T$ to write their time derivatives as total derivatives. As in \cref{sec: no slip,sec: free slip}, imposing periodicity and averaging over a period in $T$ closes the system of PDEs at this order. Without loss of generality, we assume that the period of oscillations of $p$ and $B$ is $2\pi$, and we recall the averaging operator $\avg{\cdot}$ from \cref{eq: no slip: averaging operator} as
\begin{equation}\label{eq: rollers: avg}
    \avg{a} = \frac{1}{2\pi}\int_0^{2\pi} a(T)\intd{T}\,.
\end{equation}
To compute the average of \cref{eq: rollers: first-order system} in $T$, it is instructive to consider the dependence of $f$ on its arguments explicitly. To that end, we explicitly compute the average of \cref{eq: rollers: first-order system: theta1} as
\begin{equation}\label{eq: rollers: explicit average}
    \diff{\theta_{1,0}}{t} = -\frac{3}{r^3}\left(\avg{p}\sin{\theta_{2,0}}\cos{\theta_{2,0}} + \avg{pB}\left[\sin{\theta_{1,0}}\cos{\theta_{1,0}}\left(1 - 5\cos^2{\theta_{2,0}}\right) + \cos{\theta_{2,0}}\sin{(2\theta_{1,0}-\theta_{2,0})}\right] \right)
\end{equation}
with the average of \cref{eq: rollers: first-order system: theta2} following similarly. Comparing the right-hand side of \cref{eq: rollers: explicit average} with the definition of $f$ in \cref{eq: rollers: def f}, we identify the systematically averaged governing equations with those of the original system:
\begin{subequations}\label{eq: rollers: effective eqs}
\begin{align}
    \diff{\theta_{1,0}}{t} & = f(\theta_{1,0}, \theta_{2,0}; \avg{p}, \avg{pB} / \avg{p})\,,\\
    \diff{\theta_{2,0}}{t} & = f(\theta_{2,0},\theta_{1,0}; \avg{p}, \avg{pB} / \avg{p})\,.
\end{align}
\end{subequations}
Hence, in order to understand the leading-order behaviour of the non-autonomous system of \cref{eq: rollers: non-autonomous system}, it is sufficient to explore the autonomous system
\begin{subequations}\label{eq: rollers: abstract autonomous system}
\begin{align}
    \diff{x}{t} &= f(x,y;\alpha,\beta)\,,\\
    \diff{y}{t} &= f(y,x;\alpha,\beta)\,,
\end{align}
\end{subequations}
identifying $x$ and $y$ with the leading-order solutions for $\theta_1$ and $\theta_2$, respectively. 

As noted in \cref{sec: rollers: model eqs}, the commonplace model presented by \citet{Lauga2020} makes use of constant parameters in place of $p$ and $B$ in \cref{eq: rollers: full system}, which we interpret as the \textit{a priori} averages of $p(T)$ and $B(T)$ for flow-generating particles. In symbols, when interpreted in this way, the model of \citet{Lauga2020} is equivalent to taking $(\alpha,\beta) = (\avg{p},\avg{B})$ in \cref{eq: rollers: abstract autonomous system}. The corresponding \textit{a priori}-averaged model is then given by
\begin{subequations}\label{eq: rollers: naive system}
\begin{align}
    \diff{\theta_1^a}{t} &= f(\theta_1^a,\theta_2^a;\avg{p},\avg{B})\,,\\
    \diff{\theta_2^a}{t} &= f(\theta_2^a,\theta_1^a;\avg{p},\avg{B})\,,
\end{align}
\end{subequations}
with the superscript of $a$ distinguishing the solution from that of the full system of \cref{eq: rollers: non-autonomous system}. In contrast, we have seen that the leading-order behaviour actually corresponds to taking $(\alpha,\beta) = (\avg{p}, \avg{pB}/\avg{p})$, with $\avg{pB}\neq\avg{p}\avg{B}$ in general. As we have seen throughout our analysis, this difference in parameters results directly from the employed processes of averaging: one is performed independent of the dynamical systems, whilst the other systematically determines the appropriate averaged parameters for this particular dynamical system. In the next two sections, we seek to determine if these differences in employed parameters between the systematically averaged equations and the \textit{a priori}-averaged model can result in differences in behaviour, which we establish through an elementary exploration of the autonomous dynamical system of \cref{eq: rollers: abstract autonomous system}.

\subsection{Exploring the autonomous dynamics}
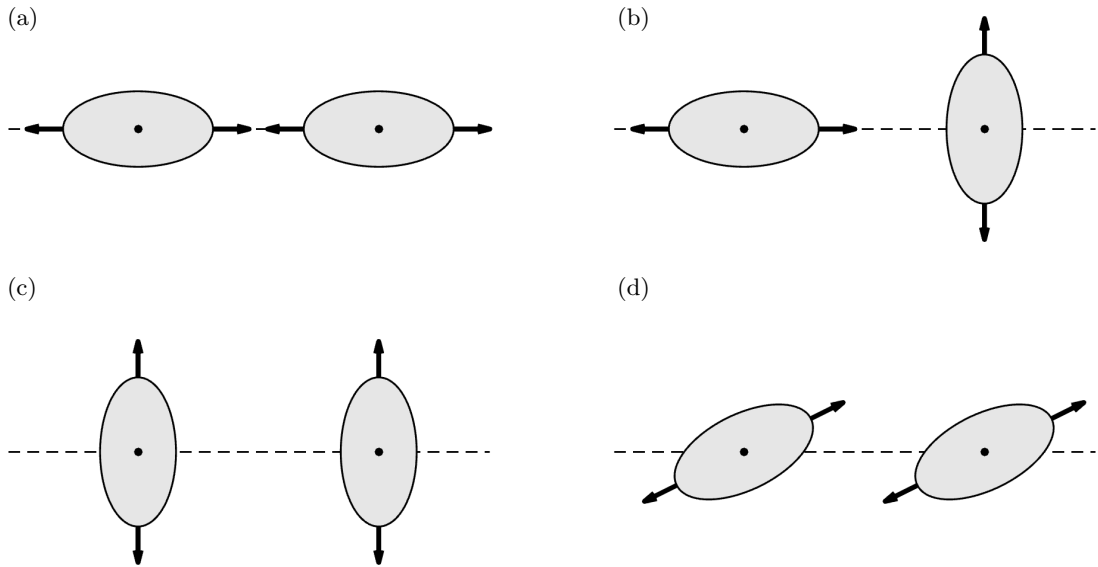
\begin{figure}
    \centering
    \begin{overpic}[width=0.8\textwidth]{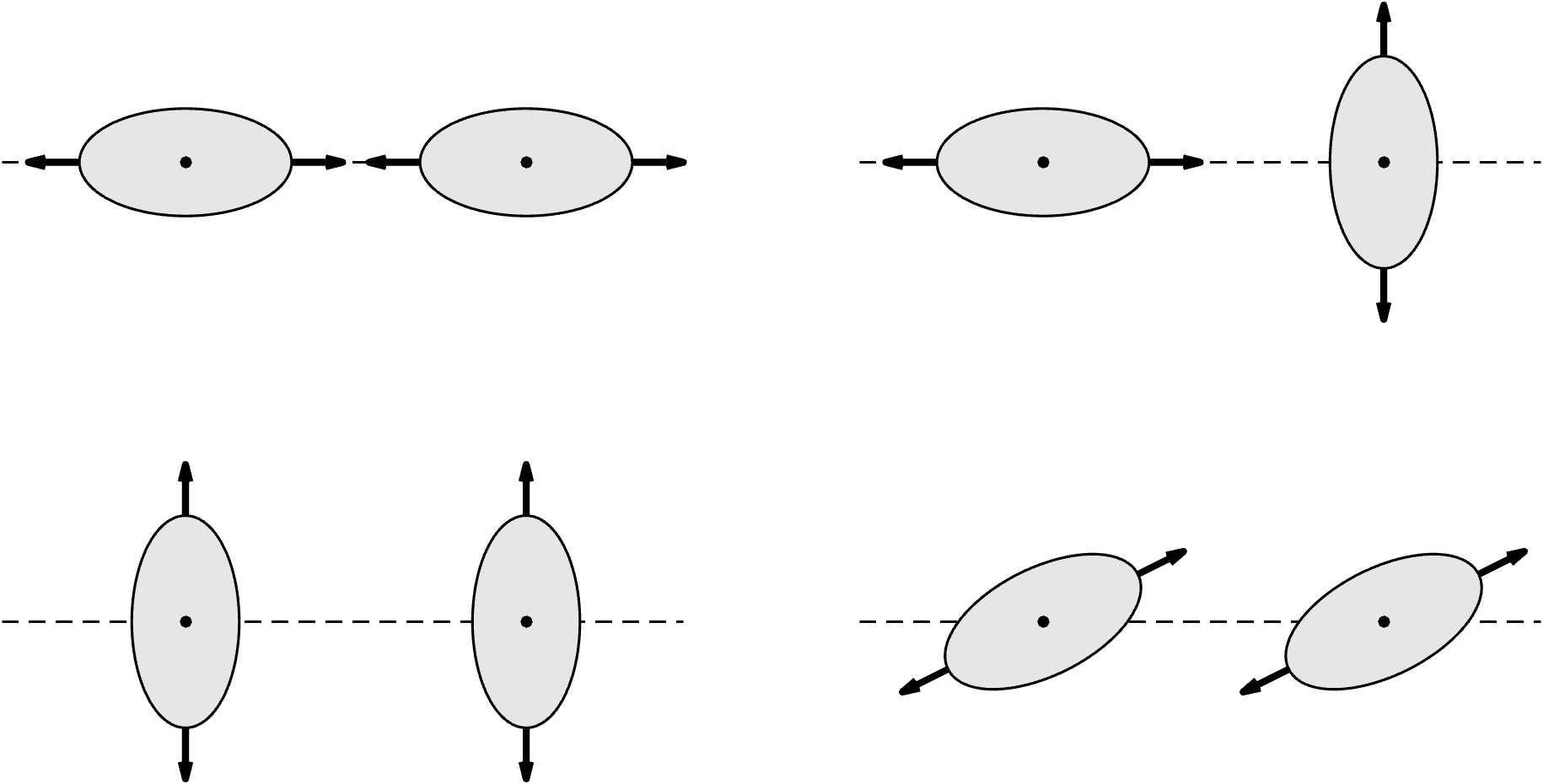}
    \put(0,50){(a)}
    \put(56,50){(b)}
    \put(0,25){(c)}
    \put(56,25){(d)}
    \end{overpic}
    \caption{Fixed points of the two-roller system. Up to symmetry and $\pi$-periodicity, the identified fixed points correspond to four distinct configurations whose stability depend on $\alpha$ and $\beta$ nontrivially. Here, we report the stability for $\alpha>0$, corresponding to a swimmer with $\avg{p}>0$ in \cref{eq: rollers: effective eqs}, which we might refer to as pusher-type particle. The dynamics of a puller-type particle, with $\alpha<0$, correspond to reversing the direction of motion from $\alpha>0$, swapping the stability of non-saddle fixed points. (a) Parallel alignment along the direction of separation is stable if $\beta<-1$ and unstable for $\beta>-1$. (b) Orthogonal alignment, with one particle pointing directly towards the other, is stable only if $\beta>0$. (c) Parallel alignment that is perpendicular to the relative displacement is only stable if $\beta<-1/2$. (d) Steady parallel alignment that is neither parallel nor perpendicular to the relative displacement is unstable when it exists, which is for $\beta < -1/2$ and $\beta>1/3$.}
    \label{fig: rollers: fixed points}
\end{figure}

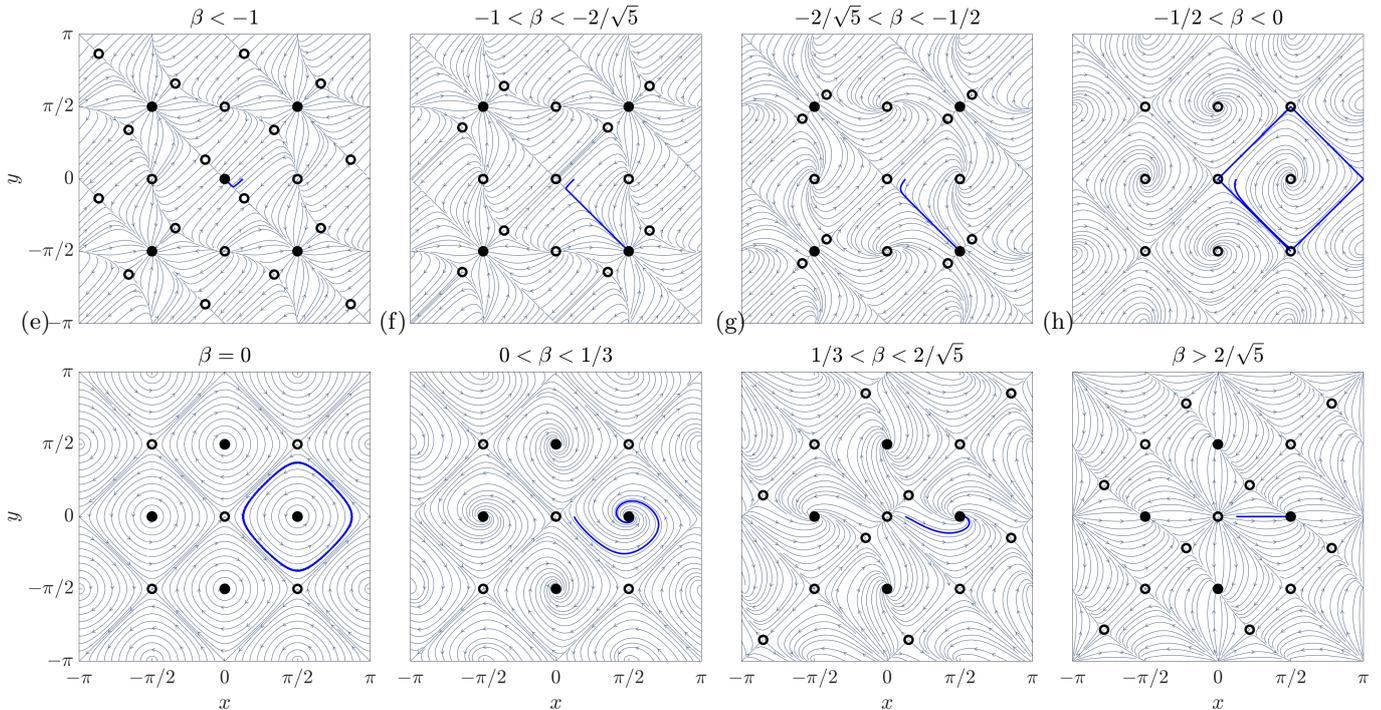
\begin{figure}
    \centering
    \begin{overpic}[permil,width=\textwidth]{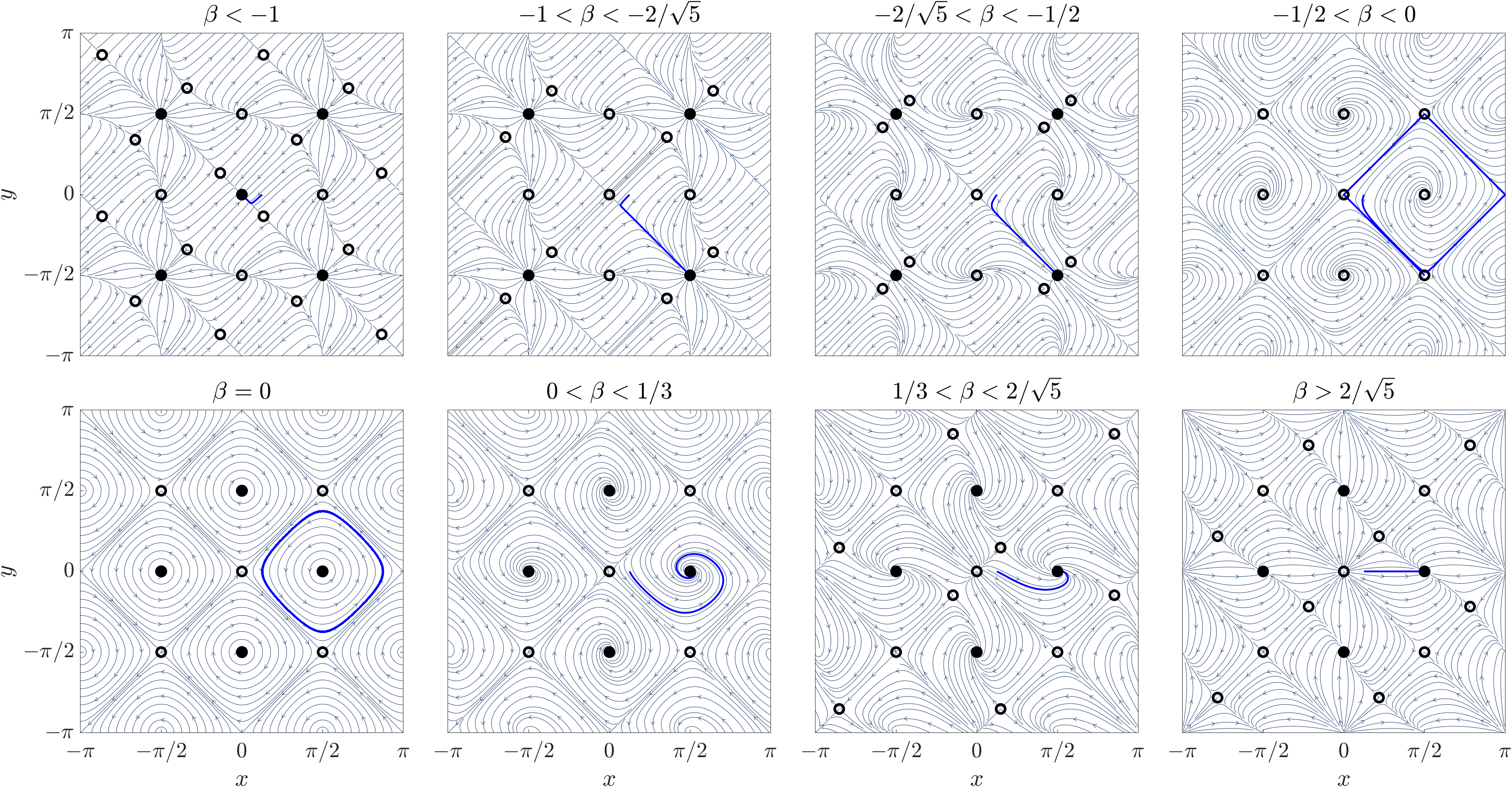}
    \put(10,557){(a)}
    \put(273,557){(b)}
    \put(520,557){(c)}
    \put(760,557){(d)}
    \put(10,282){(e)}
    \put(273,282){(f)}
    \put(520,282){(g)}
    \put(760,282){(h)}
    \end{overpic}
    \caption{Diverse phase portraits of the autonomous system of \cref{eq: rollers: abstract autonomous system}. Varying $\beta$, we illustrate the qualitatively distinct portraits of the autonomous dynamics, with stable and unstable fixed points shown as solid and empty circles, respectively. Moving from (a) to (b), we identify the transition of $(0,0)$ from a stable node to a saddle point via the coalescence of two saddles on the $x=-y$ manifold; following this transition, the parallel alignment of \cref{fig: rollers: fixed points}c is the globally attracting state. Moving from (b) to (c) and from (h) to (g) transitions stable/unstable nodes at $(0,\pi/2)$ and $(\pi/2,0)$ to spirals of the same stability. Significant qualitative changes occur through bifurcations of the saddles at $(0,0)$ and $(\pi/2,\pi/2)$, each into a node and two saddles, visible between panels (c) and (d) and panels (f) and (g). The blue trajectories in each panel, which have a common initial condition, highlight the qualitative change in behaviour that can result from changing $\beta$, potentially altering the dynamics from the almost-heteroclinic orbits of panel (d) to the distinct stable attractors of panels (c) and (f), for instance. Here, we have fixed $\alpha>0$, noting that $\alpha<0$ corresponds purely to a reversal of the dynamics.}
    \label{fig: rollers: portraits}
\end{figure}

First, we identify and classify the fixed points of \cref{eq: rollers: abstract autonomous system} in terms of $\alpha$ and $\beta$, before returning to consider the particular parameter combinations of the previous section. The dynamical system of \cref{eq: rollers: abstract autonomous system} can be explored via standard methods with relative ease, so we refrain from providing a full and detailed account of the analysis. It is worth noting that, due to the periodicity of the trigonometric functions in $f$, the forcing of the dynamics is periodic with period $\pi$ in both $x$ and $y$, so that we only need to characterise the dynamics up to multiples of $\pi$. This periodicity, along with a visual summary of the analysis that follows, is illustrated in \cref{fig: rollers: portraits,fig: rollers: fixed points}.

It is simple to identify fixed points along the manifolds $x=y$ and $x=-y$, seeking solutions of
\begin{equation}
    \frac{3\alpha}{2r^3}\sin{2x}\left(1 + \beta\left[2-5\cos^2{x}\right]\right) = 0 \quad \text{ and } \quad \frac{3\alpha}{2r^3}\sin{2x}\left(1 + \beta\cos^2{x}\right) = 0\,,
\end{equation}
respectively. For non-zero $\alpha$, these both admit solutions $x=n\pi/2$, $n\in\mathbb{Z}$, whilst the former admits the additional solutions of $1+\beta\left[2-5\cos^2{x}\right]=0$ whenever $\beta \leq -1/2$ or $\beta \geq 1/3$. There are additional steady states on the $x=-y$ manifold that satisfy $1+\beta\cos^2{x}=0$, which exist whenever $\beta\leq-1$. Further, we note the existence of additional fixed points on the manifold $x + y = \pi/2$, which again correspond to $x=n\pi/2$ for $n\in\mathbb{Z}$. Hence, the fixed points of the system are given by $(x,y)\in\{(0,0),(\pi/2,0),(0,\pi/2),(\pi/2,\pi/2)\}$, up to periodicity, in addition to solutions of $1+\beta\left[2-5\cos^2{x}\right]=0$ on the $x=y$ manifold and solutions of $1+\beta\cos^2{x}=0$ on the $x=-y$ manifold. The fixed points and their readily computed linear stabilities are summarised in \cref{tab: rollers: stab} in \cref{app: rollers: stab}, with the steady configurations interpreted in terms of the particles in \cref{fig: rollers: fixed points}. 

We illustrate the overall dynamics in various parameter regimes through phase portraits in \cref{fig: rollers: portraits}, capturing the full range of qualitatively distinct behaviours that emerge from \cref{eq: rollers: abstract autonomous system}. Noting that $\alpha$ plays only a simple role in the dynamics, with changing the sign of $\alpha$ simply reversing the direction of evolution, we fix $\alpha>0$ in \cref{fig: rollers: portraits}, focussing instead on the impact of varying $\beta$. From these portraits, it is clear that changing the value of $\beta$ can have a drastic effect on the dynamical system. For instance, $\beta$ crossing the thresholds of $-1/2$ and $1/3$ modifies the character of the phase plane through the emergence or destruction of saddle points and nodes, accompanied by qualitative changes in phase-plane trajectories. Though there are multiple further bifurcations, a notable switch in stability occurs when crossing $\beta=0$, with $\beta=0$ corresponding to an integrable system with truly closed orbits \footnote{Excluding heteroclinic trajectories.} that bifurcate into stable and unstable spirals either side of the bifurcation point.

These local bifurcations, though effecting changes in linear stability, also give rise to changes in the globally attracting dynamics of the system. This drastic alteration to the overall behaviour is illustrated via the sample trajectories highlighted in blue in \cref{fig: rollers: portraits}, which either approach closed, heteroclinic connections in \cref{fig: rollers: portraits}d or the fixed point at the centre of stable spirals in \cref{fig: rollers: portraits}f and \cref{fig: rollers: portraits}g, for instance.

\subsection{Comparing the emergent dynamics}
From the above exploration of the autonomous system of \cref{eq: rollers: abstract autonomous system}, it is clear that changes to the parameter $\beta$ can have significant qualitative impacts on the emergent dynamics. In this section, we showcase how adopting $(\alpha,\beta) = (\avg{p},\avg{B})$ in the \textit{a priori}-averaged model can give rise to predictions that are wholly different  to those of the systematically motivated parameters $(\alpha,\beta)=(\avg{p},\avg{pB}/\avg{p})$.

Before we exemplify such cases, it is worth highlighting that certain classes of $p(T)$ and $B(T)$ trivially result in $\avg{pB}/\avg{p}=\avg{B}$, so that the dynamics associated with the \textit{a priori} and systematically averaged models are identical to leading order. This equality trivially holds if at least one of $p(T)$ or $B(T)$ is constant, with $\avg{pB}=\avg{p}\avg{B}$ in this case, something which also holds for the other examples considered in this study. Hence, if the rollers can be associated with a constant dipole strength, or if their Bretherton parameters do not change in time, then we can naively average any remaining fast-time dependencies before inserting them into the model and achieve the correct leading-order behaviour. Further, if we are concerned only with the eventual configuration of the rollers, and not the details of any transient dynamics, we can identify additional $p(T)$ and $B(T)$ that can be \textit{a priori}-averaged without consequence. For instance, if $p(T)>0$ and $B(T)>0$ for all $T$, then we have $\avg{pB}/\avg{p}>0$ and $\avg{B}>0$, so that $\beta > 0$ in both cases and the configuration shown in \cref{fig: rollers: fixed points}b is globally attracting, up to symmetry, as can be deduced from \cref{fig: rollers: portraits}f-h. These particular constraints are compatible, for example, with requiring that the particle be spheroidal, always prolate, and consistently generating dipolar flow fields that can be associated with a hydrodynamic pusher.

For more general $p(T)$ and $B(T)$, however, it is clear that we cannot guarantee that the dynamics predicted by the \textit{a priori}-averaged model of \cref{eq: rollers: naive system} will be at all reminiscent of the leading-order, systematically averaged dynamics of \cref{eq: rollers: effective eqs}. Seeking a minimal example in order to highlight this general observation, we take $p(T)=8A\sin{T} + 1$ and $B(T) = (\sin{T})/4 + D$. Clearly, $\avg{p}=1$ and $\avg{B}=D$, so that the choice of $D$ uniquely determines the panel of \cref{fig: rollers: portraits} that corresponds to the dynamics of the \textit{a priori}-averaged model. However, computing $\avg{pB}/\avg{p} = A + D$ highlights that we can choose $A$ so that the systematically averaged dynamics occupy any given panel of \cref{fig: rollers: portraits}. 

\begin{figure}
    \centering
    \begin{overpic}[width=0.8\textwidth]{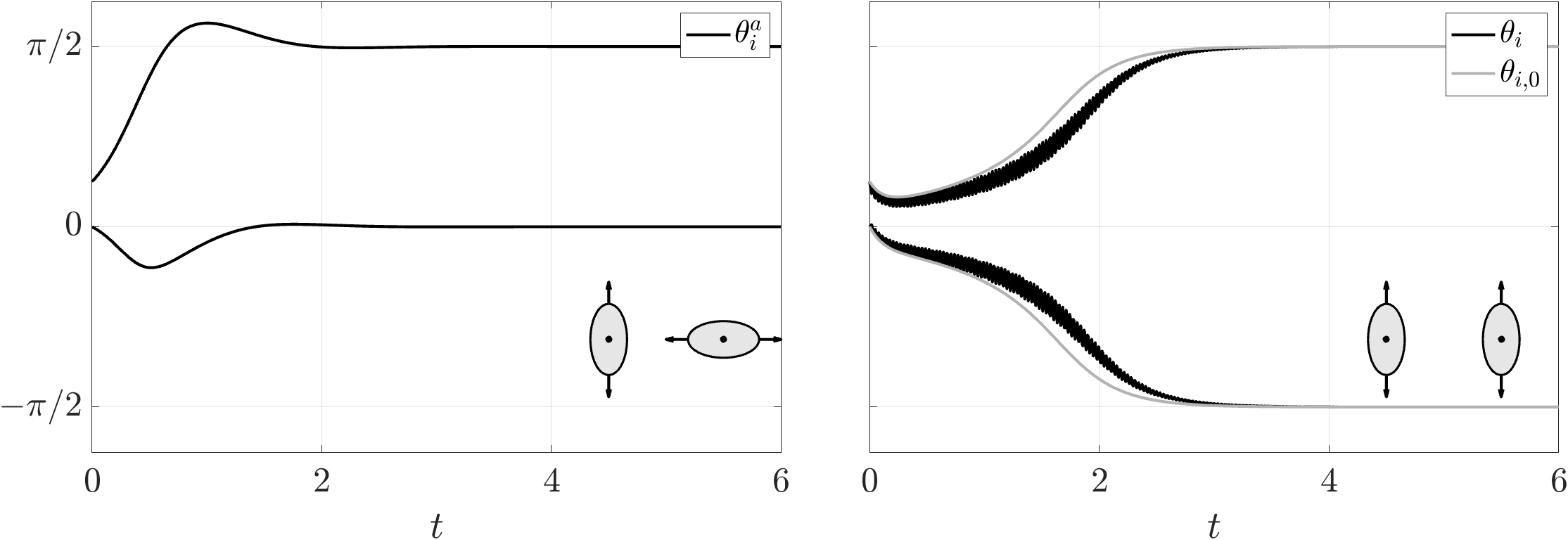}
    \put(3,36){(a)}
    \put(53,36){(b)}
    \end{overpic}
    \caption{Comparing the results of \textit{a priori} and systematic averaging. With fixed initial conditions of $(\pi/8,0)$, numerical solutions of (a) the \textit{a priori}-averaged system of \cref{eq: rollers: naive system}, (b) the systematically averaged dynamics of \cref{eq: rollers: effective eqs}, and the full system of \cref{eq: rollers: non-autonomous system} are shown as solid curves. (a) With $p(T) = -10\sin{T} + 1$ and $B(T) = (\sin{T})/4 + 1/2$, the \textit{a priori}-averaged dynamics approach the $(0,\pi/2)$ steady state, following the dynamics illustrated in \cref{fig: rollers: portraits}g, as $\beta=\avg{B}=1/2$. (b) The leading-order, systematically derived dynamics evolve to a distinct steady state, with $\theta_{1,0}=-\theta_{2,0} = \pi/2$, following the dynamics shown in \cref{fig: rollers: portraits}c, as $\beta = \avg{pB}/\avg{p} = -3/4$. The corresponding steady state configurations of the rollers are illustrated in the lower right corner of each panel, highlighting the qualitatively distinct behaviours predicted by the models. Here, we have taken $\omega=200$.}
    \label{fig: rollers: comparison}
\end{figure}

To illustrate this concretely, we take $A = -10/8$ and $D = 1/2$ and numerically solve \cref{eq: rollers: naive system,eq: rollers: effective eqs,eq: rollers: full system} from the same initial condition, taking $\omega=200$,  with the solutions shown in \cref{fig: rollers: comparison}. The \textit{a priori}-averaged system shown in \cref{fig: rollers: comparison}a corresponds to $\beta=1/2$, so follows the dynamics of \cref{fig: rollers: portraits}g, with the numerical solution approaching the $(0,\pi/2)$ steady state, in which the rollers are perpendicular to one another. In contrast, the systematically averaged dynamics of \cref{fig: rollers: comparison}b correspond to $\beta = -3/4$ and the portrait of \cref{fig: rollers: portraits}c, evolving to the parallel steady state $(\pi/2,\pi/2)$. The numerical solution to the full problem of \cref{eq: rollers: full system} is also shown in \cref{fig: rollers: comparison}b, highlighting good agreement with the asymptotic solution and evidencing the potential for disparity between the predictions of the \textit{a priori}-averaged model and the dynamics of temporally evolving bodies.

\section{Discussion and conclusions}
The use of \textit{a priori}-averaged parameters in minimal modelling approaches can be appealing, seemingly commensurate with seeking the simplest possible model of a given setting. However, through the simple examples presented in this study, we have seen how such model-agnostic averaging can result in behavioural predictions that differ qualitatively from those of models that incorporate fast-timescale parameter oscillations, which are common to many microswimmers. Hence, the key conclusion of our simple analysis is that the use of \textit{a priori}-averaged parameters can be unreliable even at the level of back-of-the-envelope calculations, with the intuition gained from such explorations potentially rendered invalid. This observation is expected to hold somewhat generically across such simple models, with our approach extending to a range of employed minimal swimmer representations.

Though one might interpret the conclusion of this study as an argument against the use of minimal models of microswimming, in fact, asymptotic analysis of these models revealed that they \emph{can} capture the leading-order dynamics, but only with appropriate parameterisation. In particular, our analysis highlights that it is the use of \textit{a priori}-averaged parameters, rather than the use of constant parameters more generally, that gives rise to inaccurate predictions. Hence, this study supports the use of minimal models in developing intuition and understanding of microswimmer systems, though only when used with systematically derived, effective parameters. 

Further, we have seen how an asymptotic analysis can show that, in certain parameter regimes, one can reliably employ \textit{a priori}-averaged parameters without qualitatively affecting the predictions of the model. However, the explorations of \cref{sec: no slip,sec: free slip,sec: rollers} revealed that such robust parameter regimes are far from being universal; on the contrary, we have seen that they depend strongly on the model in question. Hence, in general, a bespoke analysis is required for any given model in order to determine these regimes of serendipitous agreement.

The analysis presented in this study is simple, even elementary, relying on the commonplace method of multiple scales and the basic observation that the average of a product need not be the product of individual averages. Despite this simplicity, we have identified potential missteps in the use of the simplest models of microswimming, of which these authors have previously been guilty. Further, we have demonstrated how a straightforward, multi-timescale analysis can inform reliable, systematic parameterisation of minimal models, such that they recover their marked utility in the generation of intuition, basic understanding, and back-of-the-envelope predictions.

\section*{Acknowledgments}
B.J.W. is supported by the Royal Commission for the Exhibition of 1851. K.I. acknowledges JSPS-KAKENHI for Young Researchers (Grant No. 18K13456), JSPS-KAKENHI for Transformative Research Areas (Grant No. 21H05309) and JST, PRESTO, Japan (Grant No. JPMJPR1921).

\appendix
\section{Fixed points and stability of autonomous roller dynamics}\label{app: rollers: stab}
In \cref{tab: rollers: stab}, we summarise the fixed points of \cref{eq: rollers: abstract autonomous system} and their linear stability and classification. Computing these quantities is simple but cumbersome, and the dynamics are best understood through the illustrations of \cref{fig: rollers: portraits}. In \cref{tab: rollers: stab}, we have reported the stability properties for $\alpha>0$ for brevity, with the case of $\alpha<0$ leading to a precise reversal in the overall stability of the nodes and spirals of the system, with saddles remaining unstable but having their stable and unstable manifolds exchanged.
\begin{table}
    \centering
    \begin{tabular}{|c|lr|}
    \hline
        $(x,y)$ & Classification & \\
        \hline
        \multirow{3}{*}{$(0,0)$} 
            &stable node & $\beta < -1$\\
            &saddle & $-1 < \beta < 1/3$\\
            &unstable node & $\beta > 1/3$\\
        \hline
        \multirow{5}{*}{\begin{tabular}{cc}$(0,\pi/2)$\\$(\pi/2,0)$\end{tabular}}
            &unstable node & $\beta < -2/\sqrt{5}$\\
            &unstable spiral & $-2/\sqrt{5} < \beta < 0$\\
            &center & $\beta = 0$\\
            &stable spiral & $0< \beta < 2/\sqrt{5}$\\
            &stable node & $\beta > 2/\sqrt{5}$\\
        \hline
        \multirow{2}{*}{$(\pi/2,\pi/2)$}
            &stable node & $\beta < -1/2$\\
            &saddle & $\beta > -1/2$\\
        \hline
        \multirow{2}{*}{$\pm\left(\arccos{\sqrt{\frac{2\beta+1}{5\beta}}},\arccos{\sqrt{\frac{2\beta+1}{5\beta}}}\right)$}
        & \multirow{2}{*}{saddle} & $\beta < -1/2$\\
        & & $\beta > 1/3$ \\
        \hline
        $\pm\left(\arccos{\sqrt{\frac{-1}{\beta}}},-\arccos{\sqrt{\frac{-1}{\beta}}}\right)$ & saddle & $\beta < -1$\\
        \hline
    \end{tabular}
    \caption{The fixed points and linear stabilities of the autonomous system of \cref{eq: rollers: abstract autonomous system}, up to periodicity and symmetry. All stated stabilities assume that $\alpha>0$, with stability of all nodes and all spirals switching for $\alpha<0$.}
    \label{tab: rollers: stab}
\end{table}

%

\end{document}